\documentclass[aastex]{article}

\usepackage[english]{babel}
\usepackage[utf8x]{inputenc}
\usepackage[T1]{fontenc}
\usepackage{amsmath}
\usepackage{tabularx}
\usepackage{breakcites}
\usepackage{setspace}
\usepackage{gensymb}
\let\degreetwo =\degree
\let\degree = \degreethree
\usepackage{mathabx}

\let\degree = \degreetwo

\usepackage[a4paper,top=3cm,bottom=2cm,left=3cm,right=3cm,marginparwidth=1.75cm]{geometry}

\usepackage{amsmath}
\usepackage{colortbl}
\usepackage{graphicx}
\usepackage{caption}
\usepackage[colorinlistoftodos]{todonotes}
\usepackage[colorlinks=true, allcolors=blue]{hyperref}
\usepackage{authblk}
\usepackage{multicol}
\usepackage{natbib}

\title{\ Pipeline for the Detection of Serendipitous Stellar Occultations by Kuiper Belt Objects with the Colibri Fast-Photometry Array\footnote{Code for our detection pipeline and Fresnel diffraction modeling program are publicly available at github.com/ekpass/colibri}}
\author[1,2,3]{E. Pass}
\author[1,3,4]{S. Metchev}
\author[1,3]{P. Brown}
\author[5]{S. Beauchemin}

{\affil[1]{\small Department of Physics and Astronomy, University of Western Ontario, London ON, Canada}
\affil[2]{Department of Physics and Astronomy, University of Waterloo, Waterloo ON, Canada}
\affil[3]{Centre for Planetary Science and Exploration, University of Western Ontario, London ON, Canada}
\affil[4]{Department of Physics and Astronomy, Stony Brook University, Stony Brook NY, USA}
\affil[5]{Department of Computer Science, University of Western Ontario, London ON, Canada}}

\begin{document}
\maketitle

\begin{abstract}
\noindent We report results from the preliminary trials of Colibri, a dedicated fast-photometry array for the detection of small Kuiper belt objects through serendipitous stellar occultations.  Colibri's novel data processing pipeline analyzed 4000 star hours with two overlapping-field EMCCD cameras, detecting no Kuiper belt objects and one false positive occultation event in a high ecliptic latitude field.  No occultations would be expected at these latitudes, allowing these results to provide a control sample for the upcoming main Colibri campaign.  The empirical false positive rate found by the processing pipeline is consistent with the 0.002\% simulation-determined false positive rate.  We also describe Colibri's software design, kernel sets for modeling stellar occultations, and method for retrieving occultation parameters from noisy diffraction curves.  Colibri's main campaign will begin in mid-2018, operating at a 40 Hz sampling rate.

\end{abstract}
\section{Introduction}
\label{sec:intro}
\subsection{Motivation}
\label{sec:mot}
Collisional evolution models have been used to estimate the size-frequency distribution of the Kuiper belt \citep{1997Icar..125...50D, Kenyon_2004,2012MNRAS.423.1254C, Schlichting_2013}.  Both models and observation suggest that the size distribution of objects in the Kuiper belt follows a broken power law, with collisions modifying the size distribution between the power laws for large and small Kuiper belt objects \citep[KBOs;][]{Kenyon_2004, Pan_2005}.  While over a thousand KBOs larger than 15 km have been discovered, objects smaller than 10 km are too dim to be observed by direct detection, causing the distribution of small objects to be poorly constrained.

With an accurate model and a well-constrained estimate of $q$ (the power law index for the distribution of small KBOs) one could obtain insight into the material strength of these objects \citep{Pan_2005}, as well as constrain the initial planetesimal population of the Kuiper belt \citep{Schlichting_2013} and determine whether the kilometer-sized KBO population is enough to supply the Jupiter-family comets \citep{Dones_2015}.  However, without knowledge of the number of small Kuiper belt objects, $q$ cannot be well-constrained.

While small KBOs cannot be imaged directly, their presence can be inferred from serendipitous stellar occultations \citep{BAILEY_1976}.  For an occulting KBO smaller than the Fresnel scale—defined as $\sqrt[]{\lambda D/2}$ where $\lambda$ is the observation wavelength and $D$ is the distance between the occulter and observer—and for an occulted star with a projected diameter that is also small relative to this scale, an occultation event will be dominated by diffraction effects (\autoref{sec:diffr}).\footnote{For larger KBOs, the occultation approaches the geometric regime, and for angularly large stars, integration over the stellar disk results in a smoothing of the diffraction pattern.}  With a mean observing wavelength of 550 nm and a typical KBO distance of 40 AU, this characteristic Fresnel scale is 1.2 km.  Therefore, when a kilometer-sized KBO passes in front of an angularly-small star, a characteristic diffraction pattern can be observed in the star's light curve \citep{Roques_1987}.

These Fresnel diffraction patterns are observable for only a fraction of a second, but with sufficiently fast photometry, an occultation survey is able to probe the Kuiper belt for kilometer-sized objects.  In this paper, we discuss a preliminary system for such an occultation survey.

\subsection{Previous surveys}
\label{sec:prev}
A handful of occultation surveys have been performed in recent years in an attempt to characterize the size distribution of small objects in the Kuiper belt.  Early attempts by \citet{Chang_2006} and \citet{Roques_2006} each claimed to have found positive detections, but the rates found in these studies were incompatible with each other and with theoretical models \citep{Schlichting_2009}, and the calculated distances to the Roques objects would not place them within the Kuiper belt \citep{Zhang_2013}.  Subsequent studies have suggested the Chang detections were the result of systematic effects, not KBO occultations \citep{Jones_2008, Blocker_2009}.

Ground-based surveys have not yet produced any viable occultation candidates, due in part to the difficulty of performing quality fast photometry from the ground.  One such survey, a dedicated array called TAOS, did not detect any Kuiper belt objects in 292,514 star hours spread over seven years of operation \citep{Zhang_2013}.  With its sampling rate of 5 Hz, the duration of a single frame is similar to the duration of a diffraction event; the TAOS null detection is perhaps unsurprising.  A future version of the project, TAOS II, will use a sampling rate of 20 Hz to increase the likelihood of a successful detection \citep{Lehner_2014}.

\citet{Bickerton_2009} found that the Nyquist sampling rate for detection of a KBO at 40 AU in the optical is 40 Hz.  Recent ground-based efforts have focused on these high sampling rates, such as occultation surveys with Megacam on MMT \citep{Bianco_2009} and IMACS on Magellan \citep{Payne_2015}, which sampled at rates of 30 Hz and 40 Hz, respectively.  Neither have reported any detections in their 220 star hour and 10,000 star hour campaigns, although they were able to provide constraints on the surface densities of KBOs.

A potential detection of a KBO occultation was obtained by \citet{Schlichting_2009}, who analyzed 12,000 star hours of Hubble Space Telescope archival data taken at a cadence of 40 Hz.  A follow-up survey of 19,500 additional star hours yielded a second occultation candidate \citep{Schlichting_2012}.

The most stringent constraints on the small KBO population have been placed by TAOS.  TAOS established an upper limit of 3.34 to 3.82 on the index $q$ of the small-KBO power law ($\frac{dn}{dR} \propto R^{-q}$), with the size distribution of small KBOs normalized using surface densities of 38.0 $\deg^{-2}$ \citep{Fuentes_2009} and 5.4 $\deg^{-2}$ \citep{Fraser_2008} at the break in the size-frequency distribution ($R=45$ km), respectively \citep{Zhang_2013}.  Furthermore, some surveys have produced model-independent constraints at specific KBO diameters\footnote{These model-independent constraints place an upper limit on the sky surface density of KBOs of a particular size, while the TAOS results constrain $q$ for a model with a break in the size distribution at $R=45$ km, a specific surface density at the break, and a large-KBO power law index of 4.5.}, such as \citet{Bianco_2009}, \citet{Bickerton_2008}, \citet{Jones_2008} \citep[see][Figure 12]{Bianco_2009} and \citet{Schlichting_2009}.

Recent direct observations of impact craters on Pluto and Charon by the New Horizons probe have also placed constraints on the size-frequency distribution of the Kuiper belt.  Even after consideration of geological factors, these observations are suggestive of substantially lower densities of small KBOs than the maximum permitted by the upper limits established by previous occultation surveys \citep{Singer_2016}.  This low cratering rate may therefore suggest a potential unreliability in the Schlichting et al. detections \citep{Greenstreet_2015}.

\subsection{Colibri: a dedicated KBO occultation fast-photometry telescope array}
\label{sec:colibri}
Because of the rarity and brevity of sub-km Kuiper belt object occultations, a successful KBO detection campaign must optimize both sampling frequency and star hours.  The expected detection rate for KBO occultations is:

\begin{equation}
\label{eq:rate}
\mu = 2bv\Sigma\Big(\frac{180}{\pi D}\Big)^2
\end{equation}

\noindent where $b$[m] is the maximum detectable impact parameter, $v$[ms$^{-1}$] is the relative perpendicular velocity of the KBO, $\Sigma$[$\deg^{-2}$] is the sky surface density, and $D$[m] is the distance between observer and occulter \citep[see][section 3.1 for further description of these parameters]{Bickerton_2009}.  Considering a minimum detectable KBO diameter of 500 m, the time between occultations of the same star is on the scale of years to centuries, depending on the estimate used for the size-frequency distribution.

We are developing a robotic telescope array, Colibri, that will be dedicated to KBO occultation monitoring.  Colibri will consist of three 50 cm telescopes on the vertices of a triangle with sides ranging from 130 m to 270 m.  Each telescope system will image at $f/2.2$ onto a $1024 \times 1024$ electron-multiplied charge-coupled device (EMCCD) detector.  The field of view of each camera will be $0.69\degree$ on a side.  While modern $1024 \times 1024$ EMCCDs offer sampling rates that are 25 Hz at best over the full detector, by binning pixels $1\times2$ we will attain sampling rates of 40 Hz.  The limiting magnitude of individual images is expected to be 13.5 in the broad band optical at SNR=10 (similar to the $G$ band pass on Gaia).

The number of star hours obtainable from Colibri is effectively unlimited, and the 40 Hz cadence will sample $\sim$200 ms-long occultations over a number of frames.  As Colibri consists of three telescopes, candidate events can be corroborated by each independent camera and therefore a lower threshold SNR can be used to flag candidate events in each data stream, as compared to a single-camera system.

For the preliminary trials discussed in this paper, the Colibri pipeline was operated at 17 Hz using two independent EMCCD cameras, each mounted on a 75~mm $f/1.3$ video lens ($D=5.8$~cm). However, when the main campaign begins in late 2017, the array will be operating with three independent 50~cm telescope systems at 40 Hz.

\subsection{Paper Overview}
In \autoref{sec:tech}, we discuss the specifications of Colibri's preliminary observation campaign.  Our data pipeline and detection algorithm are described in \autoref{sec:soft}.  In \hyperref[sec:diffr]{section 4} we detail the theoretical expectations for the occultation light curves, and build a set of time-domain kernels for use in matched-filter detections of occultations.  \hyperref[sec:noinj]{Section 5} and \autoref{sec:retrv} discuss false positive rates and simulated event retrieval, respectively.  Results and future work are outlined in \autoref{sec:results}.

\section{Experimental Setup}
\label{sec:tech}
\subsection{EMCCD CAMO}
\label{sec:camo}
Pending the completion of the Colibri array, we performed the preliminary trials of the Colibri data pipeline (\autoref{sec:pipeline}) using the EMCCD Canadian Automated Meteor Observatory (EMCCD CAMO) array (\autoref{tab:camo}) at Elginfield Observatory in Ontario, Canada (43°11′33″N, 81°18′57″W).  As discussed in \hyperref[sec:colibri]{section 1.3}, EMCCD CAMO operates at half the cadence and with fewer, smaller telescope systems than the full Colibri array.  The purpose of these preliminary trials, therefore, is to verify the efficacy of the detection software and the plausibility of a multi-EMCCD occultation detection system at the Elginfield site, rather than to detect Kuiper belt objects.

EMCCDs' onboard gain reduces the effects of read noise, resulting in higher SNRs than typical CCDs and making them ideal for fast photometry \citep{Giltinan_2011}.  These devices have recently begun to be employed in occultation surveys, such as in the \citet{Bickerton_2008} study, the LCOGT network \citep{Bianco_2013} and the CHIMERA photometer on the Hale telescope \citep{Harding_2016}.

\noindent\begin{minipage}{\linewidth}
\centering
\bigskip
\captionof{table}{EMCCD CAMO specifications}
\begin{tabular}{l l}
  \textbf{Detectors} & 2 \\
  \textbf{Resolution} & 1024 x 1024 \\
  \textbf{Frame Rate} & 16.7 fps  \\
  \textbf{Exposure Time} & 59.91 ms \\
  \textbf{Bit Depth} & 16-bit \\
  \textbf{Pixel Scale} & 35.9"/px \\
  \textbf{Limiting Magnitude} & 8 mag at SNR=10, unfiltered $V$ band \\
  \textbf{Camera} & Nüvü HNü 1024 EMCCD \\
  \textbf{EM Gain} & 100x (set in Nüvü software) \\
  \textbf{Optics} & Navitar 75 mm f/1.3 \\
  \textbf{FOV Size} & 10.2\degree\ x 10.2\degree
\label{tab:camo}
\end{tabular}
\end{minipage}
\bigskip

\noindent\begin{minipage}{\linewidth}
\centering
\includegraphics[width=0.5\textwidth]{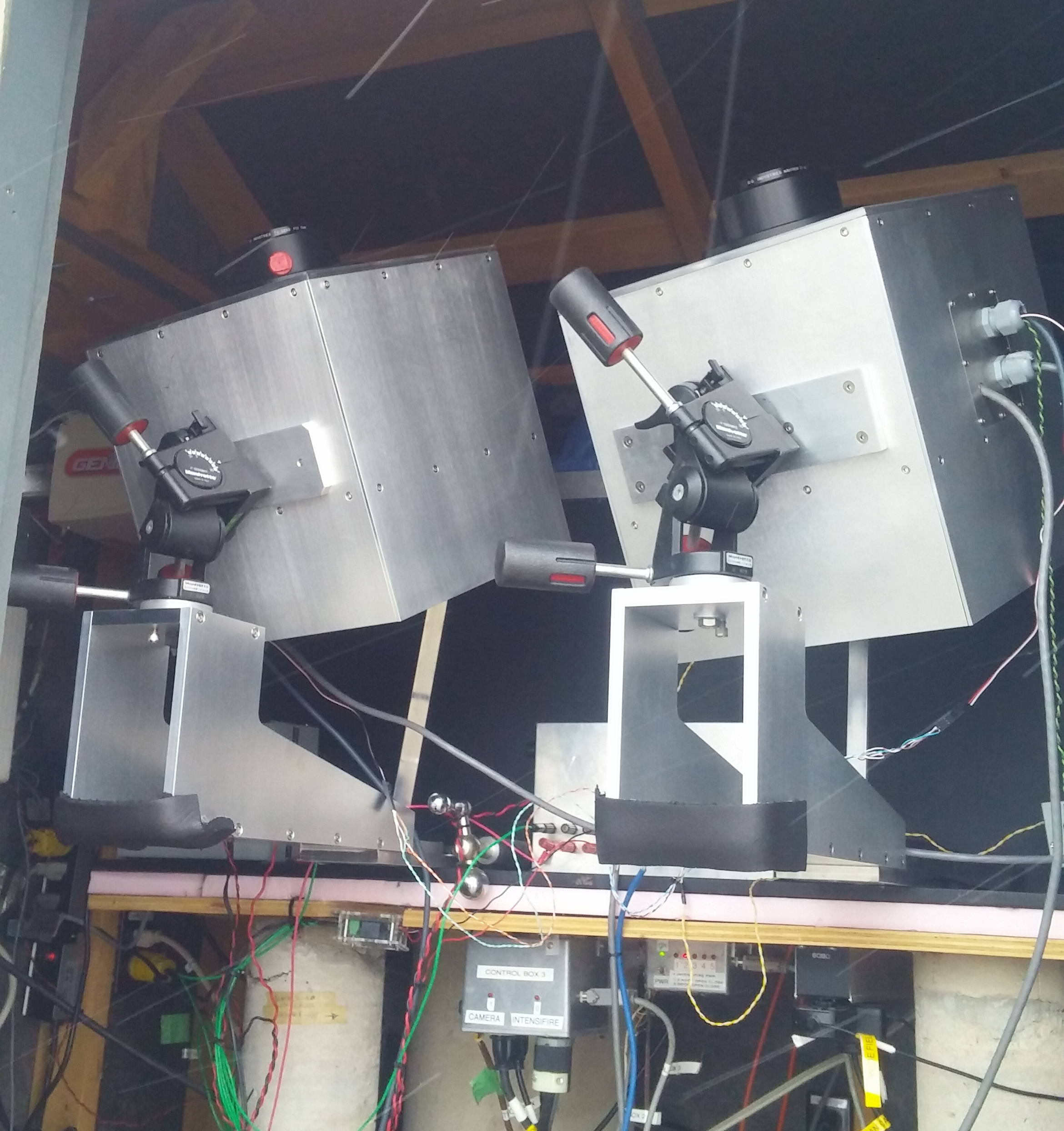}
\label{fig:EMCCD}
\captionof{figure}{The EMCCD CAMO array at the Elginfield Observatory, consisting of two Nüvü HNü EMCCD cameras with attached 75 mm lenses, pointed at a common field.}
\end{minipage}

\bigskip

The EMCCD CAMO array is autonomous, with data collection automatically beginning when skies are clear and pausing when they are cloudy.  Cloud cover is monitored by an independent camera directed at Polaris.

The Polaris monitoring module uses a Hi Cam HB310E analog NTSC camera digitized to 8 bit which has a 2\degree\ FOV.  Every five minutes, the camera does a 300 frame stack and the brightest pixels in the stacked image are selected.  Standard stellar photometry routines are used to find an apparent magnitude for Polaris. As long as the apparent magnitude is within 0.5 mag of the brightest magnitude that Polaris reaches, the skies are considered clear.  As the EMCCDs are pointed within 15\degree\ of Polaris, the sky conditions at Polaris are a very good proxy for the conditions in the EMCCD FOV.

The automation and overall design of the CAMO system has been described in detail elsewhere \citep{Weryk_2013}.

\subsection{Telescope pointing}
\label{sec:point}
A field with high densities of both angularly-small stars and KBOs is desirable for occultation surveys, although accurate photometry is difficult in overly-dense stellar fields.  Stellar field density is maximized in the galactic plane and KBO density is maximized in the plane of the Kuiper belt, which is within a few degrees of the ecliptic \citep{Brown_2004}. 

The EMCCD CAMO array's principal function is meteor detection and therefore it does not track siderally, but remains fixed in Az/Alt pointing at a zenith angle of 24.89\degree\ and an azimuthal angle of 22.86\degree.  For our observations (described in \hyperref[sec:data]{section 2.3}), this corresponds roughly to ecliptic latitudes of 45\degree\ to 82\degree\ and galactic latitudes of 37\degree\ to 58\degree.  While these telescope pointings are not optimal for KBO detection, selection of an ideal field is in progress and will be implemented for the main Colibri campaign.  This selection process involves a full characterization of the stars in the stellar field (distances, spectral types, projected angular diameters, etc.), which will allow the parameters of any candidate occultation to be thoroughly checked for consistency (see \hyperref[sec:monte]{section 6.2}).  Such a characterization process is unnecessary for the test field, as no occultation events are expected in high ecliptic latitude data.  Instead, the high ecliptic latitude data taken in these preliminary trials will serve as a control sample to the main campaign, as discussed in \autoref{sec:results}.

The overlap between the two EMCCD CAMO fields is $\sim$75\%.  As the distribution of stars that pass the detection threshold is relatively uniform across the detector, two simultaneous time series are able to be obtained for roughly 75\% of the stars detected by a single camera.

\subsection{Data acquisition}
\label{sec:data}

We used the Colibri pipeline to examine a total of 11,900 star hours.  This includes 1200 star hours observed with a single EMCCD camera on August 1st, 2016 and 10,700 star hours of observations with two EMCCD cameras between February 19th and April 1st, 2017 (\autoref{tab:dates}).  Because of the incomplete ($\sim$75\%) field overlap between the two cameras, the dual-camera observations represent a total of 4000 hours on stars with two cameras observing simultaneously.

The resulting datasets comprise 14,456 five-minute light curves from the single-camera observations on August 1st, 2016, and $\sim$48,000 five-minute light curve pairs from the dual-camera observations in early 2017.  The extracted time series in \autoref{tab:dates} represent the number of five-minute point-source time series recorded each night, the accepted time series represent the fraction of these that pass the acceptance criteria outlined in \hyperref[sec:process]{section 3.1}, and the dual-telescope time series represent the number of simultaneous light curve pairs obtained by the two EMCCD cameras.

The single-camera light curves are used in \autoref{sec:noinj} and \hyperref[sec:inj]{6.1} to determine false positive detection rates and to simulate the retrieval of occultation parameters\footnote{As discussed in \autoref{sec:noinj}, the spacing of the telescopes in the Colibri array allows for the removal of time-dependent noise effects.  As we are therefore only considering statistical noise, single-camera data can be used to generate good approximations of the false positive and event retrieval rates for the dual-camera array, provided that the systematics of the two detectors are similar (as they are in the case of EMCCD CAMO).}.  The dual-camera light curve pairs represent the actual KBO occultation experiment that we performed with the EMCCD CAMO setup, and in which we find one candidate event, likely a false positive (\hyperref[sec:nights]{section 6.3}).

\noindent\begin{minipage}{\linewidth}
\centering
\bigskip
\captionof{table}{Dates and durations of EMCCD CAMO observations}
\label{tab:dates}
\begin{tabular}{ccccc}
\hline
\hline
\multicolumn{1}{m{2cm}}
{\centering\textbf{Night}} & \multicolumn{1}{m{1.2cm}}{\centering \textbf{Hours}} & \multicolumn{1}{m{3cm}}
{\centering \textbf{Extracted \\ Time Series}} & \multicolumn{1}{m{3cm}}
{\centering \textbf{Accepted \\ Time Series}} & \multicolumn{1}{m{3cm}}
{\centering \textbf{Dual-telescope \\ Time Series}}  \\ \hline \\
1-Aug-16 & 6.67 & 49529 & 14456 & ...\\
19-Feb-17      & 7.08           & 62960                       & 16578  & 6217                    \\
22-Feb-17      & 2.25           & 16825                       & 3584 & 1344                       \\
23-Feb-17      & 1.67           & 11818                       & 2220 & 833                     \\
26-Feb-17      & 4.00           & 36365                       & 9459 & 3547                       \\
27-Feb-17      & 6.67           & 44785                       & 8886 & 3332                       \\
2-Mar-17       & 1.58           & 12136                       & 2787 & 1045                       \\
4-Mar-17       & 4.00           & 35104                       & 10404 & 3901                      \\
7-Mar-17       & 0.67           & 8006                        & 2390 & 896                       \\
16-Mar-17      & 2.83           & 23455                       & 7513 & 2817                      \\
19-Mar-17      & 0.67           & 2871                        & 511  & 192                      \\
21-Mar-17      & 4.75           & 40870                       & 10976  & 4116                    \\
22-Mar-17      & 7.83           & 66186                       & 19003 & 7126
\\
28-Mar-17 & 9.00 & 82955 & 24775 & 9291 \\
1-Apr-17 & 3.67 & 37430 & 9860 & 3698 \\ \hline

\end{tabular}
\end{minipage}
\setlength{\parindent}{15pt}

\section{Data Pipeline}
\label{sec:pipeline}
In anticipation of robotic operations by mid-2018, we have created a custom pipeline for the Colibri array that we tested with the trial data from the two EMCCD CAMO cameras.  The pipeline automatically detects when new images are taken, reduces them, and then seeks occultation detections in the time series for each star above a pre-set SNR threshold.
\label{sec:soft}
\subsection{Data processing and reduction}
\label{sec:process}
The pipeline preprocesses each five-minute data cube to remove dark and flat fielding effects, with the dark frame created from a median combination of 167 dark exposures and the flat field created from a median combination of 80 frames taken at five-minute intervals throughout one observing night.

We then extract stellar point sources using Source Extractor for Python \citep{Bertin_1996, Barbary_2016}.  The pipeline also uses the Astrometry.net API for plate solving \citep{Lang_2010} and the Astropy library within Python for various file management and convolution functions \citep{Robitaille_2013}.

The pipeline only considers time series where the net counts per frame in an $r=5$ pix (180") aperture are greater than 5000 after preprocessing and background subtraction (corresponding to an SNR$\approx$9 from empirical measurements), where the star does not leave the field of view within the first minute, and where the tracking algorithm does not otherwise fail.  For the preliminary trials, this represented approximately 1/4 of time series extracted from EMCCD CAMO data.

\subsection{Detection algorithm}
\label{sec:alg}
There are multiple techniques for numerically searching time series for KBO occultations, such as the variability index method used by \citet{Roques_2006}, the rank-probability method used by TAOS \citep{Zhang_2008}, and the template cross correlation method used by \citet{Bickerton_2008}.  We designed a novel algorithm for the Colibri array, although this structure shares some similarities with the Bickerton et al. method.

Instead of performing template matching across the entire five-minute time series (as with the Bickerton et al. algorithm), we first convolve each light curve with a Mexican hat filter, the width of which is determined by the characteristic duration of a KBO occultation at the camera sampling rate.  The duration of an occultation is $\sim$200 ms for a KBO that is small relative to the Fresnel scale and viewed at opposition\footnote{While the assumption that all objects are viewed at opposition is inaccurate given the geometry of the preliminary trials, calculations using appropriate opposition angles will be implemented for the main Colibri campaign, pending the selection of the Colibri field.} (see \hyperref[eq:lommel]{Equations 3-}\hyperref[eq:vel]{5}), which corresponds to a Mexican hat filter with a width of three frames in the 17 Hz-sampled data.  A KBO much larger than the Fresnel scale may generate a longer-duration occultation event, as the width of these occultations then becomes governed by the geometric regime rather than Fresnel optics (see \autoref{fig:width} and \citet{Nihei_2007}, Figure 3).  While our Mexican hat filter is not optimized for these larger KBOs, such events are more easily detected due to the near-complete extinction of the background star (see \citet{Nihei_2007}, Figure 5).  Indeed, our three-frame-width filter remains sensitive to objects as large as at least 5 km, as evidenced by the retrieval rates that we will discuss in \hyperref[sec:inj]{section 6.1}.

\noindent\begin{minipage}{\linewidth}
\centering
\includegraphics[width=1\textwidth]{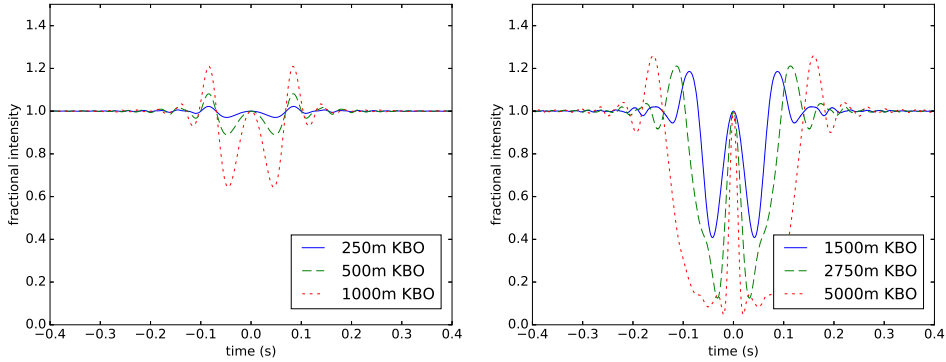}
\captionof{figure}{Theoretical occultations of a point-source star by KBOs of different sizes at 40 AU, observed in the optical, integrated over the 400 - 700 nm bandpass.  Small KBOs are shown in the left panel and large KBOs are shown in the right panel.  For objects smaller than the Fresnel scale, an increase in KBO size affects occultation depth but not width.  For larger objects, an increase in KBO diameter affects both depth and width.}
\label{fig:width}
\end{minipage}
\bigskip

The full algorithm for the Colibri pipeline consists of four phases:

\begin{enumerate}
\item The first step comprises the three-frame Mexican-hat convolution and concludes with the determination of the most significant dimming event for each five-minute convolved time series.

\item In the second step, flagged events with dips deeper than 3.75$\sigma$ below the mean of the convolved time series are allowed to proceed to the next step in the pipeline; this 3.75$\sigma$ value was empirically selected to optimize event retrieval while retaining a sufficiently low false positive rate.

\item In the third step, the candidate event is compared to a set of kernels modeling the diffraction patterns for various parameter sets (see \autoref{sec:diffr}).  With $s$=time series, $k$=kernel, $P$=Poisson noise on $s$, and $\sigma$=standard deviation of the time series\footnote{The standard deviation, $\sigma$, is influenced by the noise from EMCCD systematics and by the Poisson noise from the star's photons.  See related discussion in \hyperref[sec:inj]{section 6.1}.}, the noiseless kernel is a successful match to the noisy data when the following condition is true:
\newcommand{\dd}[1]{\mathrm{d}#1}
\begin{equation}
\label{eq:thresh}
\sum\limits_{i=1}^{len(k)} \sigma / P_i > \sum\limits_{i=1}^{len(k)}  | s_i-k_i | / P_i
\end{equation}
If \autoref{eq:thresh} evaluates true, the event proceeds to the final step in the pipeline.

\item In this step, the events flagged in step 3 are compared between the cameras.  An event which occurs at the same time in both time series is marked as an occultation candidate.
\end{enumerate}

The EMCCD cameras provide GPS-based time stamping at the image collection point, allowing the simultaneity of events to be determined with high precision.  However, the Colibri pipeline evaluates matches between the time series with a 150 ms ($\sim$3 frame) tolerance, as further testing is required to establish confidence in the EMCCD time stamping at the microsecond level.  This three-frame tolerance also acknowledges the slight differences in observed light curves that may arise from the differing geometries of the two camera systems.

The Colibri algorithm provides a heuristic determination of occultation candidates.  To quantitatively establish the confidence level of a given detection, we must benchmark the performance of this algorithm for simulated data sets.  Retrieval rates at each threshold are discussed in \hyperref[sec:inj]{section 6.1}, with false positive rates determined in \autoref{sec:noinj}.

\section{Diffraction Modeling}
\label{sec:diffr}
The diffraction pattern of an occultation can be modelled based on the radius of the KBO, the star's angular diameter, the impact parameter, the bandpass of observation, and the distance between KBO and observer (see \citet{9780521351508}, \citet{Roques_2000}, \citet{Nihei_2007}, etc).  The basics of the diffraction pattern can be calculated from Lommel functions, $U_n$, which depend only on the radius of the KBO and the Fresnel scale, itself a function of observing wavelength and distance  \citep{Roques_1987}:

\begin{equation}
\label{eq:lommel}
U_n(R, d) =\sum\limits_{k=0}^{\infty}(-1)^k * \Big(\frac{R}{d}\Big)^{n+2k}*J_{n+2k}(\pi Rd)
\end{equation}
\begin{equation}
\label{eq:occ}
I_R(d)=
\begin{cases}
      \bigg\{1+U_2^2(R, d) + U_1^2(R, d)\\
      \qquad \quad + 2\Big(U_2(R, d)\cos\big(\frac{\pi(R^2 + d^2)}{2}\big)-U_1(R,d)\sin\big(\frac{\pi(R^2 + d^2)}{2}\big)\Big)\bigg\} & d\leq R \\
      U_0^2(d, R) + U_1^2(d, R) & d\geq R \\
   \end{cases}
\end{equation}

In \autoref{eq:lommel} and \autoref{eq:occ}, $R$ is the radius of the occulter and $d$ is the distance between the line of sight and the centre of the occulter.  These values are in Fresnel scale units, which were defined in \hyperref[sec:mot]{section 1.1} as $\sqrt[]{\lambda D/2}$.  $J_{n+2k}$ is the $n+2k^{th}$ Bessel function.

$I_R(d)$ in \autoref{eq:occ} is the one-dimensional spatial intensity profile for a point-source star at a monochromatic wavelength during the occultation.  To account for a finite bandpass and stellar angular diameter, $I_R(d)$ must be convolved with the angular diameter of the stellar disk and the width of the bandpass.  Non-zero impact parameters can also be considered by changing the effective value of $d$ in the calculations of the Lommel functions.  Finally, the spatial profile can be converted to a profile in the time domain using $t=x/v_T$ and the transverse velocity  $v_T$ of the KBO, which at opposition may be approximated as \citep{Nihei_2007}:

\begin{equation}
\label{eq:vel}
v_T = v_{\Earth}\bigg(1-\sqrt[]{\frac{1.496*10^{11}}{D [\rm m]}}\bigg) [\rm ms^{-1}]
\end{equation}

\noindent\begin{minipage}{\linewidth}
\centering
\includegraphics[width=0.7\textwidth]{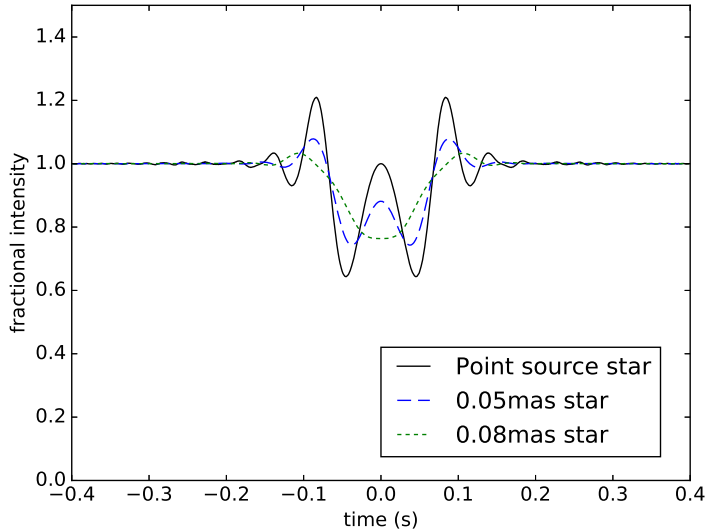}
\captionof{figure}{Occultations by a 1 km KBO at 40 AU for stars of different angular diameters, $\theta$.  Diffraction features begin to disappear with realistic sampling (\hyperref[fig:rates]{Figures 4}\hyperref[fig:shift2]{-6}) and noise levels (\autoref{fig:det}) as stellar angular diameter is increased.}
\label{fig:angdi}
\end{minipage}
\bigskip

\autoref{fig:angdi} shows the diffraction patterns that would be observed if the exposure times were infinitesimally short.  Since every camera has some finite exposure time, to generate the pattern that would be observed by these devices one must integrate over exposure length.

\noindent\begin{minipage}{\linewidth}
\centering
\includegraphics[width=0.7\textwidth]{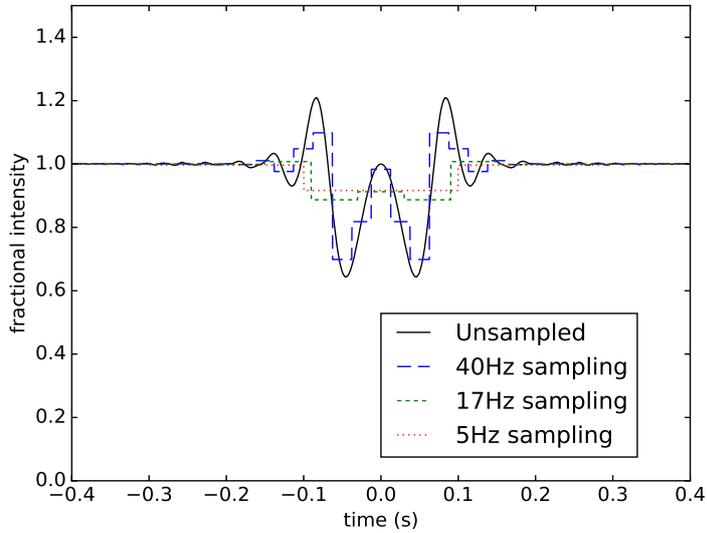}
\captionof{figure}{The same occultation of a point-source star as shown in \autoref{fig:angdi}, integrated at different sampling rates.  The event spans a single frame with 5 Hz sampling and three frames when sampled at 17 Hz.  Diffraction features are only visible in the 40 Hz-sampled curve.}
\label{fig:rates}
\end{minipage}
\bigskip

As the patterns in \autoref{fig:rates} are dependent on the time at which the exposures were centred, we implement an additional parameter in our simulations:  an arbitrary shift adjustment with respect to the middle of the exposure.

\noindent\begin{minipage}{\linewidth}
\centering
\includegraphics[width=0.7\textwidth]{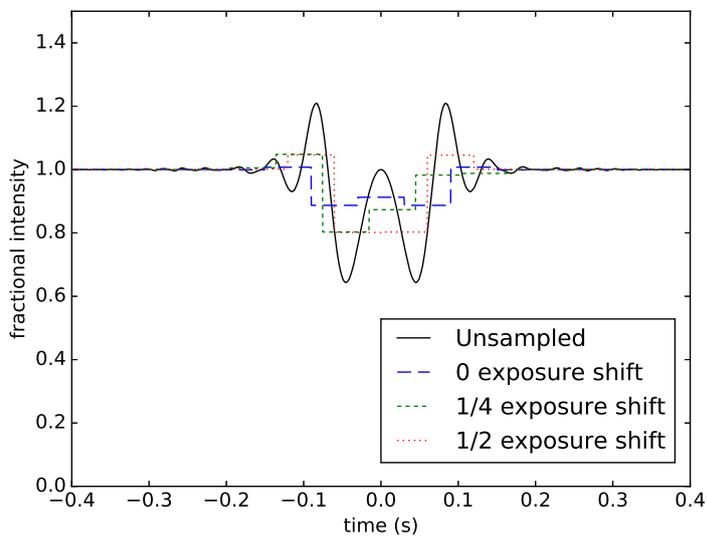}
\captionof{figure}{The same occultation of a point-source star as shown in \autoref{fig:angdi},  sampled at 17 Hz and with varying shift adjustments.  Due to finite exposure times, changes in shift adjustment can lead to large differences in the light curve.}
\label{fig:shift}
\end{minipage}
\bigskip

\noindent\begin{minipage}{\linewidth}
\centering
\includegraphics[width=0.7\textwidth]{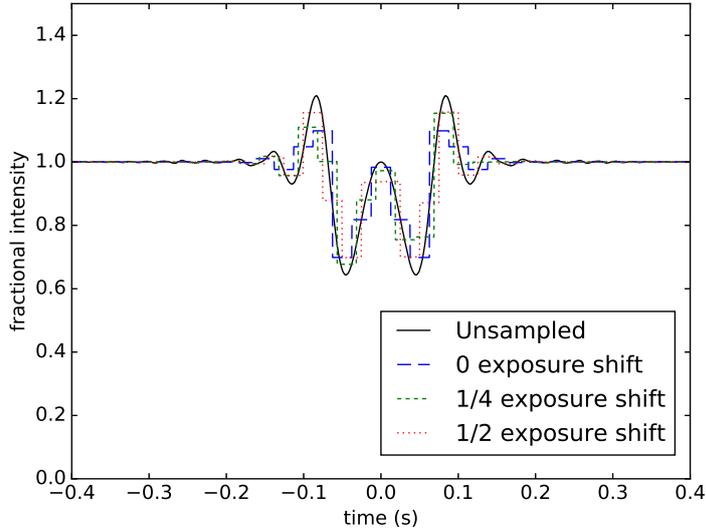}
\captionof{figure}{The same occultation of a point-source star as shown in \autoref{fig:angdi}, sampled at 40 Hz and with varying shift adjustments.  When the light curve is critically sampled, shift adjustments have little effect on the curve shape.  This is another advantage of a fast sampling rate.}
\label{fig:shift2}
\end{minipage}
\bigskip

We generated a set of 54 kernels for the preliminary tests of the Colibri pipeline.  These span a parameter space that includes KBO radii of 750 m, 1375 m, and 2500 m, stellar angular diameters of 0.01 mas, 0.03 mas, and 0.08 mas, shift adjustments of 0, 0.25, and 0.5 fractions of an exposure, and impact parameters of 0 and the KBO radius.

This kernel set serves multiple purposes.  Firstly, the noiseless kernels can be used  in the kernel matching step of the Colibri pipeline (\hyperref[sec:alg]{section 3.2}).  Secondly, noise can be added to these kernels to simulate data taken during an occultation event, and these noisy kernels can then be used to characterize retrieval rates (\hyperref[sec:inj]{section 6.1}).  There is some degeneracy in the parameter space defined by these kernels, which we will explore in \hyperref[sec:monte]{section 6.2}.

\section{False Positives}
\label{sec:noinj}

Because of the rarity of serendipitous stellar occultations, false positives can easily outnumber true events if the algorithm's false positive rate is not kept sufficiently low.  Characterizing the false positive rate of the Colibri pipeline is therefore critical to determining whether the algorithm is capable of distinguishing true occultation events from the noise.

False positive detections can be statistical in nature, or they can be the result of physical phenomena, such as scintillation events in the upper atmosphere.  The 130 m to 270 m separations between the telescopes of the Colibri array have been selected such that scintillation events will not appear simultaneously in the three data streams (see \citet{Lehner_2009}, who determined that a 6 m separation between 50 cm telescopes is sufficient to eliminate time-correlated noise; the separation between the 50 cm telescopes of the Colibri array is over twenty times larger).  By eliminating time-correlated noise effects in this way, we may focus our investigation on statistical false positives.  We therefore implement a randomization of the time stamps of the EMCCD CAMO light curves in order to obtain a large data set to estimate Colibri’s false positive rate; such randomization removes any time-correlated noise, but retains the statistical false positive events that are currently under consideration.

The 14,456 single-camera EMMCD CAMO light curves from August 1st, 2016 were used to generate 780,624 randomized time series pairs. After processing the randomized time series with the Colibri pipeline, 18.7\% had a dip deeper than 3.75$\sigma$ below the mean after the three-frame Mexican-hat convolution, 11.1\% were successfully matched to a template kernel, and there was a simultaneous detection in 16 time series pairs.  The final false positive rate of the Colibri pipeline is therefore 0.002\%.

This false positive rate is compatible with the results of our on-sky observations (\hyperref[sec:data]{section 2.3}).  In 48,000 tests of the Colibri algorithm on data from the two EMCCD CAMO systems, only one false positive was detected. This is fully consistent with the expected 0.002\% false positive rate from our simulations.  However, a single false positive would also be consistent with a much broader range of expectation rates, ranging from 0.01 to 7.5 events in 48,000 tests at the 99.5\% confidence level.  With a practically unlimited number of star-hours for the future Colibri experiment, such large uncertainties in the performance will be eliminated.

Further elimination of false positives is possible with subsequent analysis of the time series flagged by the Colibri pipeline.  Despite identifying that an event occurs simultaneously between the cameras, Colibri makes no effort to ensure that the dip is comparable in size and shape.  A process for verifying that two curves correspond to an occultation with consistent parameters is described in \hyperref[sec:monte]{section 6.2}.

\section{Occultation Detection}
\label{sec:retrv}
In this section, we discuss the Colibri pipeline's ability to successfully detect serendipitous stellar occultations.  In \hyperref[sec:inj]{section 6.1}, noisy occultation events are injected into EMCCD CAMO light curves to characterize Colibri's retrieval rate and range of detectable parameters.  In \hyperref[sec:monte]{section 6.2}, we use a Monte Carlo simulation to determine the extent to which we can retrieve occultation parameters from noisy light curves.  From these simulations, we can establish confidence levels for our parameter estimates.  \hyperref[sec:monte]{Section 6.3} discusses the event detected by the preliminary trials of the Colibri pipeline, which we classify as a false positive through Monte Carlo simulations.

\subsection{Retrieval of injected events}
\label{sec:inj}

We will consider the noise in our time series in terms of two components:  the Poisson-distributed shot noise resulting from the photons measured from the star ($\sigma_P$), and all other noise ($\sigma_O$).  We can therefore write:

\begin{equation}
\label{eq:noise1}
\sigma = \sqrt[]{\sigma_P^2 + \sigma_O^2}
\end{equation}

\noindent where $\sigma$ is the standard deviation of our light curve.

We can also calculate $\sigma_P$:

\begin{equation}
\label{eq:noise2}
\sigma_P = {\sqrt[]{\frac{N }{G}}*G}
\end{equation}

In \autoref{eq:noise2}, $N$ is the mean counts in the time series and $G$ is the EMCCD detector gain ($G$=100 for EMCCD CAMO).  Using \autoref{eq:noise1} and \autoref{eq:noise2}, we can determine $\sigma_O$ from measurements of $N$ and $\sigma$.

During an occultation event, there is a decreased number of photons arriving at the detector from the star. To include noise in the kernels of our kernel set, we must consider this variability.

For each kernel, we replace each point with a random variate drawn from a Poisson distribution with $\lambda = \frac{N*k}{G}$, where $k$ is the fractional flux at that point in the kernel.  This variate is then multiplied by $G$, as our light curves are in units of counts.  Finally, we apply $\sigma_O$ using a random Gaussian distribution.  In \autoref{fig:statmodel}, we illustrate that our two-component noise model closely reproduces the noise effects in our observed light curves.

\noindent\begin{minipage}{\linewidth}
\centering
\includegraphics[width=0.7\textwidth]{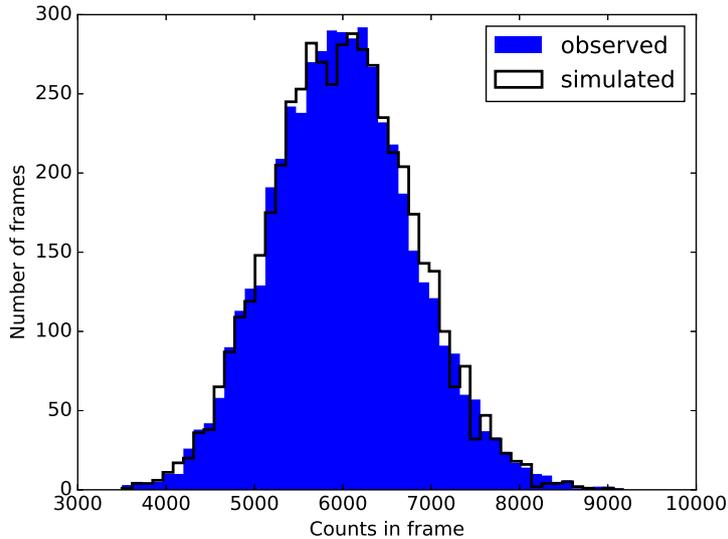}
\captionof{figure}{The filled histogram shows the observed distribution of counts for a representative light curve from our sample.  The black line shows the distribution derived from our statistical model; each point in the simulated light curve is a random variate drawn from a Gaussian distribution with $\mu=0$ and $\sigma = \sigma_O$, added to a random variate (multiplied by $G$) drawn from a Poisson distribution with $\lambda = N/G$.}
\label{fig:statmodel}
\end{minipage}
\bigskip

To investigate the Colibri pipeline's event retrieval rates, noise-convolved versions of each kernel in the kernel set were injected into 14,456 five-minute time series.  These series correspond to the stars that appeared in EMCCD CAMO data taken on August 1st, 2016 and passed the acceptance criteria outlined in \hyperref[sec:process]{section 3.1}.

After injection, the Colibri pipeline attempted to retrieve these simulated events.  \autoref{tab:inj} benchmarks the performance of this method at each of the four steps described in \hyperref[sec:alg]{section 3.2}.

The first five columns of \autoref{tab:inj} identify each kernel by its occultation parameters.  As the first step in occultation detection, the pipeline selects the most significant flux dip in each five-minute time series after the Mexican-hat convolution (\hyperref[sec:alg]{section 3.2}).  However, depending on the parameters of the simulation kernel, this dip does not always correspond to the simulated event.  The fraction under the “Correct Dip Detected” column in \autoref{tab:inj} shows how frequently the injected event was the one identified as the most probable occultation candidate in the time series.  The entries in the subsequent columns show the cumulative fractions of originally simulated events that continue to be recovered after each discriminating step in the detection pipeline.
\bigskip

\noindent\begin{minipage}{\linewidth}
\centering
\captionof{table}{Retrieval rates of kernels injected into EMCCD CAMO light curves}
\label{tab:inj}
\begin{tabular}{ccccccccc}
\hline
\hline
\multicolumn{1}{m{1.2cm}}{\centering\scriptsize \textbf{Kernel} \scriptsize}
& \multicolumn{1}{m{1.2cm}}{\centering \scriptsize Object Diameter \\ (m) \scriptsize}
& \multicolumn{1}{m{1.3cm}}{\centering \scriptsize Stellar Diameter \\ (mas) \scriptsize}
& \multicolumn{1}{m{1.3cm}}{\centering \scriptsize Impact Parameter \\ (m) \scriptsize}
& \multicolumn{1}{m{1.5cm}}{\centering \scriptsize Shift Adjustment \\ (frame) \scriptsize} & \multicolumn{1}{m{1.5cm}}{\centering \scriptsize Correct Dip Detected \\ (\%) \scriptsize}
& \multicolumn{1}{m{1.3cm}}{\centering \scriptsize Pass Dip Threshold \\ (\%) \scriptsize}
& \multicolumn{1}{m{1.5cm}}{\centering \scriptsize Pass Kernel Match \\ (\%) \scriptsize}
& \multicolumn{1}{m{1.25cm}}{\centering \scriptsize Pass Dual Detection \\ (\%) \scriptsize} \\ \hline \\
\textbf{1}	&	1500	&	0.01	&	0	&	0	&	32	&	22	&	17	&	4.4	\\
\textbf{2}	&	1500	&	0.01	&	750	&	0	&	61	&	51	&	42	&	22	\\
\textbf{3}	&	1500	&	0.01	&	0	&	0.25	&	34	&	25	&	20	&	5.5	\\
\textbf{4}	&	1500	&	0.01	&	750	&	0.25	&	64	&	54	&	46	&	25	\\
\textbf{5}	&	1500	&	0.01	&	0	&	0.5	&	37	&	27	&	22	&	7.0	\\
\textbf{6}	&	1500	&	0.01	&	750	&	0.5	&	67	&	58	&	49	&	28	\\
\textbf{7}	&	2750	&	0.01	&	0	&	0	&	99.28	&	98.9	&	82	&	69	\\
\textbf{8}	&	2750	&	0.01	&	1375	&	0	&	89	&	85	&	72	&	54	\\
\textbf{9}	&	2750	&	0.01	&	0	&	0.25	&	99.31	&	98.9	&	84	&	71	\\
\textbf{10}	&	2750	&	0.01	&	1375	&	0.25	&	88	&	83	&	71	&	54	\\
\textbf{11}	&	2750	&	0.01	&	0	&	0.5	&	99.22	&	98.9	&	87	&	77	\\
\textbf{12}	&	2750	&	0.01	&	1375	&	0.5	&	87	&	81	&	68	&	50	\\
\textbf{13}	&	5000	&	0.01	&	0	&	0	&	100	&	100	&	94.7	&	90.2	\\
\textbf{14}	&	5000	&	0.01	&	2500	&	0	&	99.34	&	99	&	84	&	72	\\
\textbf{15}	&	5000	&	0.01	&	0	&	0.25	&	100	&	100	&	95	&	90.5	\\
\textbf{16}	&	5000	&	0.01	&	2500	&	0.25	&	99.44	&	99.14	&	86	&	75	\\
\textbf{17}	&	5000	&	0.01	&	0	&	0.5	&	100	&	100	&	94.7	&	90.2	\\
\textbf{18}	&	5000	&	0.01	&	2500	&	0.5	&	99.45	&	99.17	&	85	&	74	\\
\textbf{19}	&	1500	&	0.03	&	0	&	0	&	36	&	26	&	20	&	6.3	\\
\textbf{20}	&	1500	&	0.03	&	750	&	0	&	55	&	45	&	37	&	17	\\
\textbf{21}	&	1500	&	0.03	&	0	&	0.25	&	38	&	28	&	23	&	7.4	\\
\textbf{22}	&	1500	&	0.03	&	750	&	0.25	&	57	&	47	&	39	&	19	\\
\textbf{23}	&	1500	&	0.03	&	0	&	0.5	&	40	&	30	&	25	&	8.6	\\
\textbf{24}	&	1500	&	0.03	&	750	&	0.5	&	59	&	49	&	42	&	21	\\
\textbf{25}	&	2750	&	0.03	&	0	&	0	&	99.65	&	99.46	&	85	&	73	\\
\textbf{26}	&	2750	&	0.03	&	1375	&	0	&	89	&	84	&	71	&	54	\\
\textbf{27}	&	2750	&	0.03	&	0	&	0.25	&	99.57	&	99.42	&	86	&	75	\\
\textbf{28}	&	2750	&	0.03	&	1375	&	0.25	&	87	&	82	&	71	&	53	\\
\textbf{29}	&	2750	&	0.03	&	0	&	0.5	&	99.6	&	99.39	&	87	&	77	\\
\textbf{30}	&	2750	&	0.03	&	1375	&	0.5	&	86	&	81	&	69	&	51	\\
\textbf{31}	&	5000	&	0.03	&	0	&	0	&	100	&	100	&	95.7	&	91.9	\\
\textbf{32}	&	5000	&	0.03	&	2500	&	0	&	99.38	&	99	&	86	&	76	\\
\textbf{33}	&	5000	&	0.03	&	0	&	0.25	&	100	&	100	&	95.8	&	91.8	\\
\textbf{34}	&	5000	&	0.03	&	2500	&	0.25	&	99.38	&	99	&	87	&	77	\\
\textbf{35}	&	5000	&	0.03	&	0	&	0.5	&	100	&	100	&	95.3	&	91.1	\\
\textbf{36}	&	5000	&	0.03	&	2500	&	0.5	&	99.37	&	99.07	&	87	&	76	\\
\textbf{37}	&	1500	&	0.08	&	0	&	0	&	47	&	36	&	30	&	12	\\
\textbf{38}	&	1500	&	0.08	&	750	&	0	&	24	&	16	&	13	&	2.5	\\
\textbf{39}	&	1500	&	0.08	&	0	&	0.25	&	47	&	37	&	31	&	13	\\
\textbf{40}	&	1500	&	0.08	&	750	&	0.25	&	24	&	16	&	13	&	2.6	\\
\textbf{41}	&	1500	&	0.08	&	0	&	0.5	&	47	&	36	&	31	&	12	\\
\textbf{42}	&	1500	&	0.08	&	750	&	0.5	&	24	&	16	&	13	&	2.6	\\
\textbf{43}	&	2750	&	0.08	&	0	&	0	&	99.7	&	99.61	&	86	&	75	\\
\textbf{44}	&	2750	&	0.08	&	1375	&	0	&	77	&	69	&	59	&	39	\\
\textbf{45}	&	2750	&	0.08	&	0	&	0.25	&	99.67	&	99.57	&	87	&	77	\\
\textbf{46}	&	2750	&	0.08	&	1375	&	0.25	&	78	&	70	&	61	&	41	\\
\textbf{47}	&	2750	&	0.08	&	0	&	0.5	&	99.68	&	99.62	&	86	&	75	\\
\textbf{48}	&	2750	&	0.08	&	1375	&	0.5	&	77	&	70	&	60	&	40	\\
\textbf{49}	&	5000	&	0.08	&	0	&	0	&	100	&	100	&	95	&	90.4	\\
\textbf{50}	&	5000	&	0.08	&	2500	&	0	&	97.5	&	96.4	&	84	&	72	\\
\textbf{51}	&	5000	&	0.08	&	0	&	0.25	&	100	&	100	&	95.1	&	91.1	\\
\textbf{52}	&	5000	&	0.08	&	2500	&	0.25	&	97.7	&	96.5	&	85	&	74	\\
\textbf{53}	&	5000	&	0.08	&	0	&	0.5	&	100	&	100	&	94.9	&	90.5	\\
\textbf{54}	&	5000	&	0.08	&	2500	&	0.5	&	97.7	&	96.5	&	84	&	72	\\
\hline
\end{tabular}
\end{minipage}

\setlength{\parindent}{15pt}
\bigskip

Retrieval rates from the final dual-detection step of our event retrieval simulation range from $\sim$2.5\% to $\sim$92\% (\autoref{tab:inj}), illustrating a significant dependence on the occultation parameters, including KBO size, stellar angular diameter, and impact parameter.  Kernels 38, 40, and 42, the templates with the lowest successful retrieval rates in \autoref{tab:inj}, model small KBOs (1.5 km) occulting angularly-large stars (0.08 mas) with large impact parameters (750 m).

\autoref{tab:inj} shows that these kernels perform particularly poorly at the first threshold:  identification as a candidate for further analysis using the Mexican-hat convolution.  Due to scintillation and other noise, these $\sim$25\% dips often cannot be distinguished from non-occultation fluctuations.  As the survey's sensitivity to these type of events is already so low, we calculate that removing these kernels from the kernel set for a 17 Hz-sampled, 5.8 cm aperture survey would reduce the false positive rate by a greater factor than it would reduce occultation sensitivity.

As discussed in \hyperref[sec:colibri]{section 1.3}, the expected occultation rate varies greatly depending on the index $q$ of the small-object power law.  This rate also has a significant dependence on KBO size.  Using \autoref{eq:rate} and related equations from \citet{Bickerton_2009} and assuming $q=$1.9 \citep{Fraser_2008}, we calculate that the number of occultations by 1500 m to 2750 m objects is approximately 1.7x the number of occultations by 2750 m to 5000 m objects.\footnote{Alternatively, we can calculate this value as 5.5x using the $q=$3.8 upper limit established by TAOS.  As the \citet{Singer_2016} result suggests that $q$ is substantially lower than the TAOS upper limit, we will proceed using $q=1.9$ for this investigation.  However, it is important to note that the true value of $q$ may vary from this estimate.}

If the nine kernels with final retrieval rates less than 10\% were removed from the kernel set, the pipeline's final false positive rate would decrease fourfold, from 0.002\% to 0.0005\%.  Above, we calculated that approximately two thirds of the occultations that may be detected by this kernel set correspond to objects smaller than 2750 m.  Therefore, removing these 1500 m kernels from the kernel set could result in a threefold decrease in KBO sensitivity at worst.  Realistically, the sensitivity decrease would be less severe:  only a fraction of kernels sensitive to 1500 m to 2750 m objects would be removed, and the removed kernels have much lower retrieval rates (and therefore are sensitive to a smaller fraction of the events they are theoretically capable of detecting) than the other kernels.

Nevertheless, it would be beneficial to lower the minimal detectable diameter of KBOs and improve retrieval rates at these small diameters rather than simply removing the low-sensitivity kernels.  The number of occultations by objects between 500 m and 1500m is 2.6x the number of occultations by objects between 1500 m and 5000 m:  lowering the minimal detectable diameter would greatly improve the likelihood and increase the frequency of occultation detections.  We aim to decrease the minimal detectable diameter of KBOs with the main Colibri campaign, which will utilize much larger (50 cm) telescopes.

In addition, the main campaign will use 40 Hz sampling as opposed to the 17 Hz sampling of these preliminary trials.  At a 40 Hz sampling rate, characteristic diffraction patterns in the time series are a strong differentiator of occultation events from random noise or scintillation (\autoref{fig:shift2}).  From the data in \autoref{tab:inj} and the false positive rates described in \autoref{sec:noinj}, the kernel match step may not appear to be a significant benefit to the detection pipeline, as diffraction patterns are not clearly visible in the data with 17 Hz sampling (\autoref{fig:rates} and \autoref{fig:shift}).  We anticipate a substantial improvement in retrieval rates at the kernel matching step with the 40 Hz main Colibri campaign.

The kernel set will need to be expanded and reoptimized prior to main campaign data collection to ensure that the 40 Hz templates with visible diffraction patterns are sensitive to the full spectrum of parameter values, including the lower KBO diameters available to the Colibri array.  Based upon previous studies, a set of 300 or more kernels may be required to ensure sensitivity throughout the parameter space for a survey conducted at a 40 Hz sampling rate with a noise level of 2-4\% \citep{Bickerton_2008}.

\subsection{Parameter estimation}
\label{sec:monte}
Using a Monte Carlo simulation based on the diffraction modeler described in \autoref{sec:diffr}, we are able to estimate the parameters that generate a given occultation pattern.  This process allows KBO characteristics to be determined in the event of a successful detection.

Additionally, this method may be used to filter out false positive candidates that were not removed by the main Colibri pipeline.  By constraining the occultation parameters for the dips detected by each camera, discrepancies between the two curves can be analyzed quantitatively.  Disparate curves can then be flagged as false positives.

\bigskip
\noindent\begin{minipage}{\linewidth}
\centering
\includegraphics[width=0.7\textwidth]{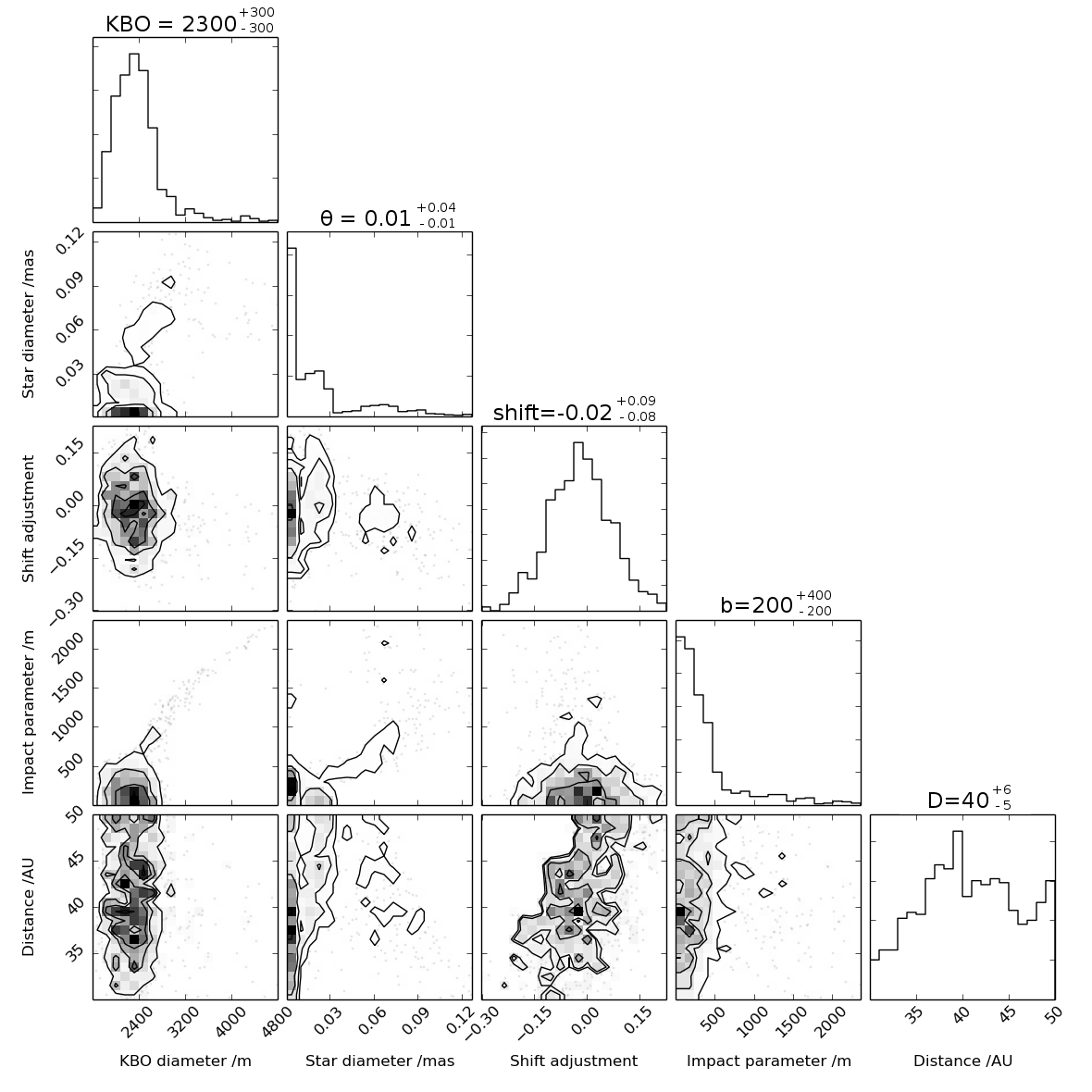}
\captionof{figure}{Parameter estimates for 1000 runs of the Monte Carlo simulation program.\protect\footnotemark\  Stellar angular diameter has been allowed to vary freely for these simulations.  However, following Data Release 2 of the Gaia mission, we will use actual stellar diameter estimates.}
\label{fig:corner}
\end{minipage}
\footnotetext{Plotting was performed using the corner.py Python package \protect\citep{Foreman_Mackey_2016}.}
\bigskip

\noindent\begin{minipage}{\linewidth}
\centering
\includegraphics[width=0.7\textwidth]{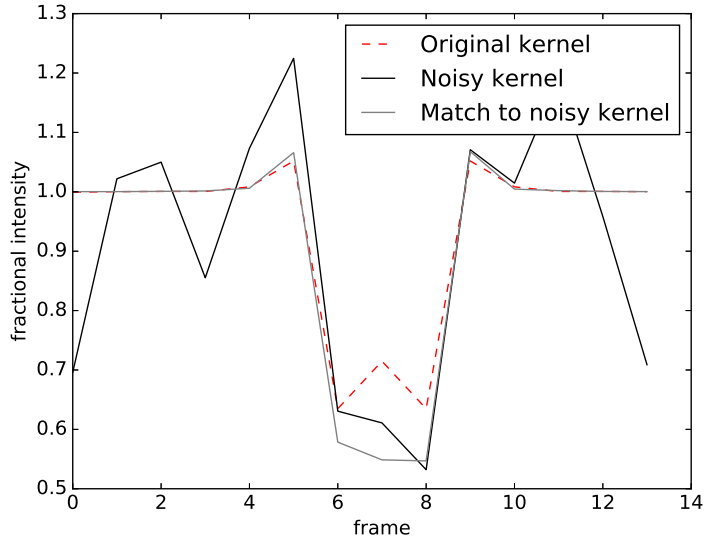}
\captionof{figure}{The light curve defined by the best fit parameters from \autoref{fig:corner} for the noise-convolved kernel (KBO diameter=2300 m, $b$=200 m, $\theta$=0.01 mas, $D$=40 AU, shift=-0.02).}
\label{fig:noisy}
\end{minipage}
\bigskip

\autoref{fig:corner} shows the parameter estimates for an occultation in a high-noise time series (SNR=6).  The original input parameters were a 2000 m KBO, a point source star (0 mas stellar angular diameter), 0 m impact parameter, 0 shift adjustment, and a distance of 40 AU, with the kernel corresponding to these parameters displayed in \autoref{fig:noisy}.  Despite this time series exhibiting even higher noise levels than allowed by Colibri's acceptance criteria, the histograms in \autoref{fig:corner} show that for each parameter except distance, the simulation results are clustered around the input value.  Distance cannot be well constrained with 17 Hz data.

Some degeneracy can be observed between KBO size and impact parameter.  This is unsurprising, as a larger KBO deepens the occultation curve while an increased impact parameter shallows it.  In the case of \autoref{fig:corner}, this may account for the increase in estimated KBO size and impact parameter as compared to the simulation input.  This degeneracy is therefore an effect to consider when evaluating the results of this method on non-simulated data.

\subsection{Results of the Colibri occultation search}
\label{sec:nights}

From the 4000 star hours of dual-camera data obtained in the present experiment (\hyperref[sec:data]{section 2.3}), the Colibri pipeline flagged one candidate occultation of a 6.72 V-magnitude star (SNR=7) on February 26th.

\noindent\begin{minipage}{\linewidth}
\makebox[\textwidth][c]{\includegraphics[width=1.3\textwidth]{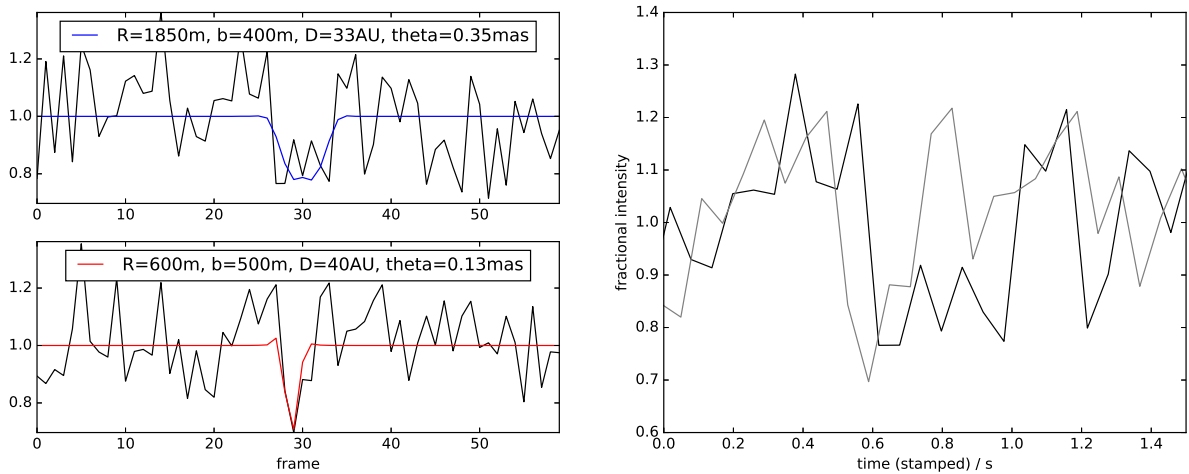}}
\captionof{figure}{Light curves observed by each camera for the candidate occultation, with inconsistent parameter estimates derived through Monte Carlo simulation (see \hyperref[sec:monte]{section 6.2}).  If microsecond precision of the EMCCD CAMO time stamps is assumed, the candidate event appears offset between the time series, as shown in the right panel.}
\label{fig:det}
\end{minipage}
\bigskip

Using the Monte Carlo parameter retrieval method described in \hyperref[sec:monte]{section 6.2}, we found that the occultation in one time series corresponded to a KBO of radius 1800 m $\pm$ 500 m, while we determined a KBO radius of 600 m $\pm$ 200 m using the second time series.  The two estimates of the radius are marginally inconsistent: a t-test results in a p value of 0.026, so the estimates are inconsistent at the 97.4\% level.  For this reason, we classify the candidate as a likely false positive.  The impact parameter estimate is the same within uncertainties for the two curves, which suggests that the size/impact parameter degeneracy is not contributing to the discrepant KBO size estimates.

\section{Summary and Future Work}
\label{sec:results}
After analysis of 4000 star hours recorded simultaneously by two independent cameras, we report no Kuiper belt occultations in the preliminary trials of the Colibri pipeline.  This is not unexpected:  our measurements were taken at ecliptic latitudes of 45\degree\ to 82\degree.  As the sky surface density of KBOs is near zero at such high ecliptic latitudes \citep[Figure 10]{Volk_2017}, data collected in this regime cannot be used to constrain KBO surface density, but instead provides a control for other measurements taken near the ecliptic.  For example, event rate measurements at $|\beta|>20\degree$ have been used by previous surveys \citep[e.g.,][]{Schlichting_2009} as zero-detection control samples to establish false positive rates.

In processing $\sim$48,000 five-minute light curve pairs (4000 star hours), one false positive was detected by the pipeline.  This is consistent with the 0.002\% false positive rate found in our simulations.  These preliminary trials therefore establish a control sample for the future Colibri campaign.

As discussed in \citet{Bianco_2009}, a dedicated array such as TAOS is required to establish the most stringent constraints on the small-KBO population.  On the other hand, fast-photometry campaigns are required for sizes, distances, and impact parameters to be determined from a detected occultation (a process described in \hyperref[sec:monte]{section 6.2}).  While the latter type of survey represents the majority of recent occultation work, these surveys do not use dedicated arrays.  Dedicated fast-photometry occultation surveys like Colibri and the upcoming TAOS II are therefore ideally suited to measure the population properties of small KBOs.

\bigskip
\noindent\textbf{Acknowledgements}

\medskip
\noindent We thank the anonymous referee for a very insightful and helpful assessment of this manuscript.  We also express our gratitude to the University of Western Ontario Meteor Physics Group for their technical support with the EMCCD CAMO system, particularly Jason Gill and Zbyszek Krzeminski.  In addition, we thank Wayne Hocking for valuable discussions regarding scintillation effects and Richard Bloch for providing comments on this manuscript.

This work has been supported in part by a Centre for Planetary Science and Exploration Interdisciplinary Undergraduate Research Award and by a Natural Sciences and Engineering Research Council of Canada Undergraduate Student Research Award.
\bigskip

\noindent\begin{minipage}{\linewidth}
\bibliographystyle{aasjournal}
\begin{small}
\setlength{\bibsep}{0.0pt}
\bibliography{kbopaper}
\end{small}
\end{minipage}

\end{document}